\pdfoutput=1
\documentclass{aa}  
\usepackage{graphicx}
\usepackage{natbib}
\usepackage{color}
\usepackage{txfonts}
\def\xmmu{XMMU J054134.7-682550}
\def\xmmtwomass{2MASS $05413431-6825484$ }
\def\rxte{\emph{RXTE }}
\def\xmm{\emph{XMM-Newton }}

\begin{document}
   \title{Pulsed thermal emission from the accreting pulsar \\  \xmmu \thanks{ 
                  Based on observations obtained with XMM-Newton, an ESA science mission with instruments and contributions directly funded by 
                  ESA Member States and NASA}}
   \subtitle{}

   \author{A. Manousakis\inst{1,2}\and R. Walter\inst{1,2}\and M. Audard\inst{1,2}\and T. Lanz\inst{3} }

   \offprints{A. Manousakis}

   \institute{ISDC Data Center for Astrophysics, Chemin d'Ecogia 16, CH-1290 Versoix, Switzerland \\
              \email{Antonios.Manousakis@unige.ch}
         \and Observatoire de Gen\`eve, Universit\'e de Gen\`eve,  Chemin des Maillettes 51, CH-1290 Versoix, Switzerland  
         \and Department of Astronomy, University of Maryland College Park, MD 20742-2421, USA}

   \date{Received October 3, 2008; accepted February 2, 2009}

\titlerunning{Pulsed thermal emission from XMMU J054134.7-682550}
\authorrunning{A. Manousakis et al.}

 
  \abstract
   {}
   {Soft X-ray excesses have been detected in several Be/X-ray binaries and interpreted as the signature of hard X-ray reprocessing in the inner accretion disk. The system \xmmu, located in the LMC, featured a giant Type II outburst in August 2007. The geometry of this system can be understood by studying the response of the soft excess emission to the hard X-ray pulses.}
   {We have analyzed series of simultaneous observations obtained with XMM-Newton/EPIC-MOS and RXTE/PCA in order to derive spectral and temporal characteristics of the system, before, during and after the giant outburst. Spectral fits were performed and a timing analysis has been carried out. Spectral variability, spin period evolution and energy dependent pulse shapes are analysed.}
   {The outburst $({\rm L}_{\rm X}= 3\times 10^{38} {\rm erg/s}\approx  {\rm L}_{\rm EDD})$ spectrum could be modeled successfully using a cutoff powerlaw, a cold disk emission, a hot  blackbody, and a cyclotron absorption line.  The magnetic field and magnetospheric radius could be constrained. The thickness of the  inner accretion disk is broadened to a width of 75~km. The hot blackbody component features sinusoidal modulations indicating that the bulk of the hard X-ray emission is emitted preferentially along the magnetic equator. The spin period of the pulsar decreased very significantly during the outburst. This is consistent with 
   a variety of neutron star equations of state and indicates a very high accretion rate. 
   }
   {}


   \maketitle
%

\section{Introduction}

Be/X-ray binaries consist of a neutron star orbiting a Be star, which is defined as a non-supergiant B-type star whose spectrum shows (or showed, at some time)
one or more Balmer lines in emission.
Be/X-Ray binaries display X-ray pulsations, a signature of the strong magnetic field ($B\sim10^{12}\,$G) of the neutron star orbiting a massive star companion.
Most known Be/X-Ray binaries undergo outbursts in which their X-ray luminosity suddenly increases by a factor of $\sim 10 - 10^{4}$ with respect to the quiescence level.

They can feature two types of outbursts:
Type I (or normal)
X-ray outbursts of moderate intensity (${\rm L}_{X}\sim  10^{36}$ erg s$^{-1}$)
occuring during the periastron passage of the neutron star and
Type II (or giant)
X-ray  outbursts of higher intensity (${\rm L}_{X} \sim 10^{37-38}$ erg s$^{-1}$) lasting for several
weeks or even months. Generally, Type II outbursts start shortly after
periastron passage, but do not show any other correlation with orbital
parameters \citep{Finger_Prince_97}.
A small fraction of Be/X-ray binares are persistent sources (the prototype being X-Per), with a low luminosity ${\rm L}_{X}\sim10^{34}$ erg s$^{-1}$ at an almost constant emitting level \citep{Reig_Roche_99}.

During giant outbursts, the spin period of the neutron star has been observed to decrease (neutron star spin-up), indicating that angular momentum is transfered
from the accreted material to the neutron star, through an accretion disk \citep{Finger_et_al_99,Wilson_et_al_03}.

\citet{Corbet_86} has shown that Be/X-Ray binaries fall into a narrow area in the ${\rm P_{spin}-P_{orb}}$ diagram. This correlation has been interpreted as a result of the rotation of the neutron star at 
the equilibrium velocity between the spin-up, and the spin-down led by centrifugal effects 
of the strong magnetic field \citep{Waters_et_al_89}.

The X-ray spectra of Be/X-Ray binaries are very close to those of accreting pulsars, although these depend on the physical conditions close to the neutron star.
The spectra can be characterized by cutoff powerlaws. In a few systems with low interstellar absorption, there is 
evidence for soft excesses at low energies, often modeled as blackbody components \citep{White_et_al_83,Hickox_et_al_04,Paul_et_al_02,Endo_et_al_00}.
\citet{Hickox_et_al_04} showed that the soft excesses observed in luminous X-ray sources (${\rm L_{X}}>10^{38}\,$erg s$^{-1}$) can only be explained by reprocessing of hard X-rays 
by optically thick material, near the inner edge of the accretion disk. Many, if not all, bright sources with low absorption have shown this feature.

\xmmu, located in the Large Magellanic Cloud -  LMC,   has been proposed as a likely HMXB by \citet{Shtykovskiy_et_al_05}.
\citet{Palmer_et_al_07} found \xmmu$\,$ in a flaring state during a routine scan of the 
Swift-BAT data on August 3, 2007 at a level of $\approx 50$ mCrab. Subsequent RXTE observations  
on August 9, 2007 revealed X-ray pulsations (${\rm P_{s}}\approx 61.601 \pm 0.017$s) and cyclotron features at 10 keV \citep{Markwardt_et_al_07}. 
Assuming that  this source is a Be system having a (giant) type II outburst, \citet{Markwardt_et_al_07} estimated an orbital period of about 80 days (within a factor of $\sim$2) 
based on the Corbet diagram. The average PCA (Proportional Counter Array on board RXTE)  spectrum could be fit with a cut-off power law, with 
photon index 0.47 and an e-folding cut-off energy of 16 keV \citep{Markwardt_et_al_07}.

In this paper we present XMM-Newton and RXTE data revealing a soft pulsed thermal emission. 
In section 2, we present the observations and data reduction; in section 3, the spectral and timing analysis. 
In section 4, we discuss the source reprocessing geometry and summarize our results. 

\begin{table*}[!]
\caption{XMMU J$054134.7-682550$ observing runs. The table lists  the initial time of the exposure, the exposure time, and 
the number of source counts. In the case of  XMM-Newton, the number of counts is the sum obtained for both 
EPIC-MOS detectors.}
\label{table:obsrun}
\centering
\begin{tabular}{c | c c r | c c r }
\hline\hline
Observation & XMM-Newton      &                &                          & RXTE                 &                &                          \\
            & Start           & Exposure (ks)  & Counts                   & Start                & Exposure (ks)  & Counts      \\
\hline
1 & 2007-07-06T23:32:48 & 10.4                 &    15241                 &  -                   &                &            \\
2 & 2007-08-21T15:12:12 & 7.5                  &    33184                 & 2007-08-21T19:07:28  & 3.5            &  192650          \\
3 & 2007-10-05T23:50:15 & 17.0                 &    22576                 & 2007-10-05T21:57:52  & 1.6            &  24537          \\
4 & 2007-11-24T22:06:15 & 19.3                 &    4332                  & -                    &                &            \\
\hline
\hline
\end{tabular}
\end{table*}


\section{Observations and data reduction}

\subsection{Observations}

Figure \ref{table:obsrun} displays the long-term RXTE/ASM light curve of \xmmu, featuring an
outburst with a duration of about 50 days \citep{Markwardt_et_al_07}.
The four observing runs used in this work were carried out using 
the PCA on board the \emph{Rossi X-ray Timing Experiment} (RXTE) and EPIC-MOS on board \emph{XMM-Newton} (XMM) \citep{Jansen_et_al_01}.  
Table \ref{table:obsrun} lists the start time, exposure, and number of source counts for each observation. 
For observation $2$ and $3$ the XMM and RXTE exposures were performed quasi-simultaneously. RXTE/PCA data were not available for observations $1$ and $4$.

\begin{figure}
\centering
\includegraphics[width=0.45\textwidth, bb=0 0 700 600]{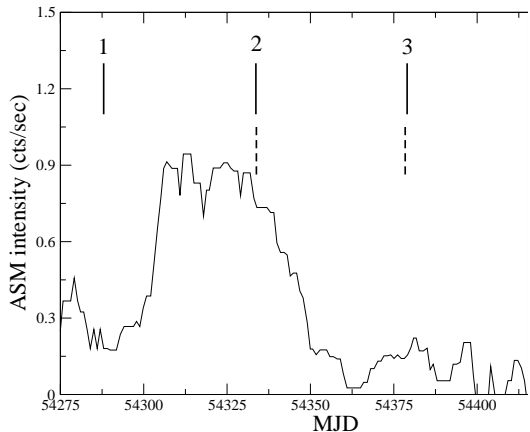}
\caption{RXTE/All-Sky-Monitor long term light-curve (2 days average) of \xmmu. The pointed  observations of  XMM-Newton and RXTE are indicated by 
solid and dashed vertical lines, respectively.}
\label{fig:outburst_rxte}
\end{figure}

\subsection{Data reduction}
\subsubsection*{\emph{Rossi X-ray Timing Experiment} (RXTE)} 
We used data from  RXTE obtained in late August 2007 and 
early October 2007 (Table \ref{table:obsrun}).   
The PCA \citep{Jahoda_et_al_06} instrument consists of 5 identical multianode Proportional Counter Units (PCU), operating in the $2-60$ keV energy band, with an effective 
area of approximately $6500$ cm$^{2}$ and a 1 degree FWHM field of view. 
PCA spectra and light-curves were extracted using standard FTOOLS (HEASOFT\footnote{http://heasarc.gsfc.nasa.gov/lheasoft/} version 6.3.1).
Data were accumulated from Standard-2 mode\footnote{http://heasarc.gsfc.nasa.gov/docs/xte/xtegof.html}.  The Bright Background model 
($>40 {\rm \hspace{0.1cm} cts \hspace{0.1cm} s  }^{-1}{\rm PCU}^{-1}$) was used for observation 2 at the peak of the outburst, and the Faint Background model was used for observation 3. 
We applied good time intervals (GTI) with an elevation greater than 10 degrees and a pointing offset angle 
less than 0.02 degrees. Response matrices were created using the tool \verb=pcarsp=. 

\subsubsection*{ \xmm}
Serendipitous observations of \xmmu$\,$  were performed by EPIC-MOS on board \xmm for four epochs between July and November 2007
(obsID: 0500860301 - 0500860601, PI: Lanz)
The Science Analysis Software (SAS) version 7.1.0 was used to produce event lists for the EPIC-MOS[12] \citep{Turner_et_al_01}   
instrument running \verb=emchain=. EPIC MOS was operated in small window mode for the pointed object, i.e. Cal 83. All the rest of the CCDs were operated in full frame 
mode providing a time resolution of 2.6 s. 
Significant pile-up effects were identified in obsevation 2 and 3 and were reduced by selecting data within an annular ring around the piled-up area, although
out of time events were not identified. The source was outside the field of view of the other XMM-Newton instruments.

\section{Data analysis}

\subsection{Outburst spectrum}
\label{subsection:spectral_an}

\begin{figure}[!b]
\centering
\includegraphics[width=0.45\textwidth,angle=0]{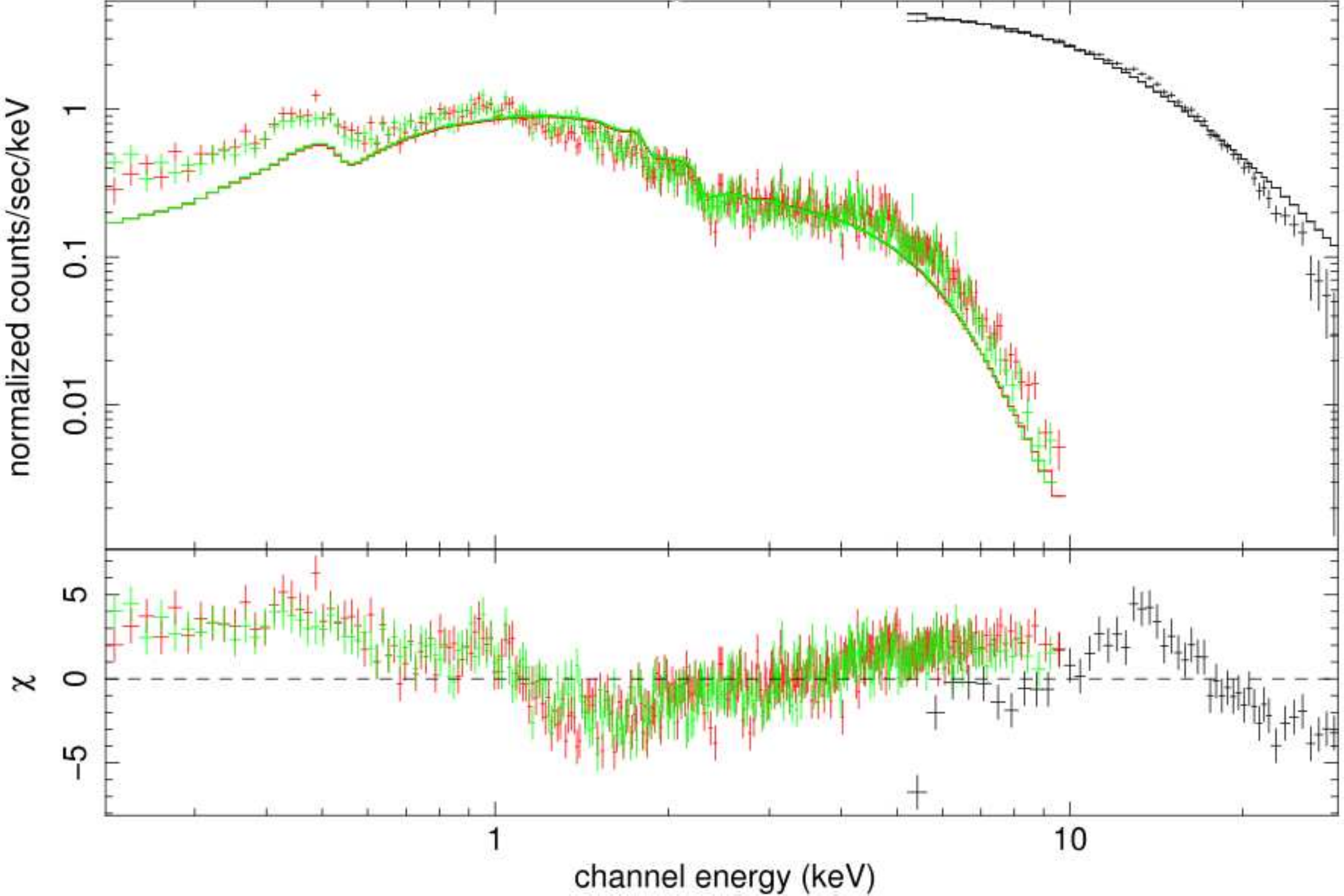}
\caption{Outburst spectrum obtained using  EPIC/MOS1+2 and RXTE/PCA, fit with a simple absorbed cut-off powerlow model (const*wabs*cutoff).}
\label{fig:sepc:ds2:test1}
\end{figure}

The spectral analysis was performed using the XSPEC\footnote{http://xspec.gsfc.nasa.gov/} package version 11.3.2 \citep{Arnaud_et_al_07}. 
For the outburst spectrum quasi-simultaneous observations were available, i.e., observation 2.
We selected data from EPIC/MOS1+2 and RXTE/PCA covering the energy band $0.2-10.0$ keV and $5.0-30.0$ keV, respectively.
To fit the spectrum,  we initially used a very simple model, made of three components: photoelectric absorption, a powerlaw with high energy exponential cutoff, 
and a cross-calibration factor, i.e., \verb=const*wabs*cutoff=. 
The model failed to fit the data, providing  $\chi_{\nu}^{2}\sim 3$ (Fig \ref{fig:sepc:ds2:test1}).
The residuals (Fig. \ref{fig:sepc:ds2:test1}, lower panel) indicate a clear soft excess below $1$ keV and spectral features in the $7-20$ keV energy range.  

We added a fourth spectral component in order to account for the soft excess: either a blackbody (\verb=bbody=) or a disk black body (\verb=diskbb=).   
This additional spectral component improves the fit very significantly, providing  $\chi_{\nu}^{2} \sim 1.2$ for \verb=diskBB= and $\chi_{\nu}^{2} \sim 1.5$ 
for \verb=bbody=, and a cross-calibration factor of $\sim 0.9$ in both cases.

The best fit column density  ${\rm N_{H}}\sim 2.1\times10^{20}$ cm$^{-2}$ is, however, less than the galactic value
in the source direction ${\rm N_{H}^{{\rm gal}}}\sim 6.32\times 10^{20} {\rm cm}^{-2}$ \citep{DL_90}. Fixing the column density to the galactic value,
and using only one thermal model, an additional excess is observed below $\sim 0.5$ keV, indicating that one thermal model
is not adequate to represent the data. The model can be further improved by adding both disk and blackbody spectral components. 
To better constrain the parameters, we decided to fix the disk's inner radius to the radius of the magnetosphere,
$R_{m} \sim  1.3\times 10^{8}$ cm, inferred from the cyclotron line. 
The disk component turned out to the softer (and cooler) than the single blackbody emitting region.

\begin{figure}
\centering
\includegraphics[width=0.5\textwidth,angle=0]{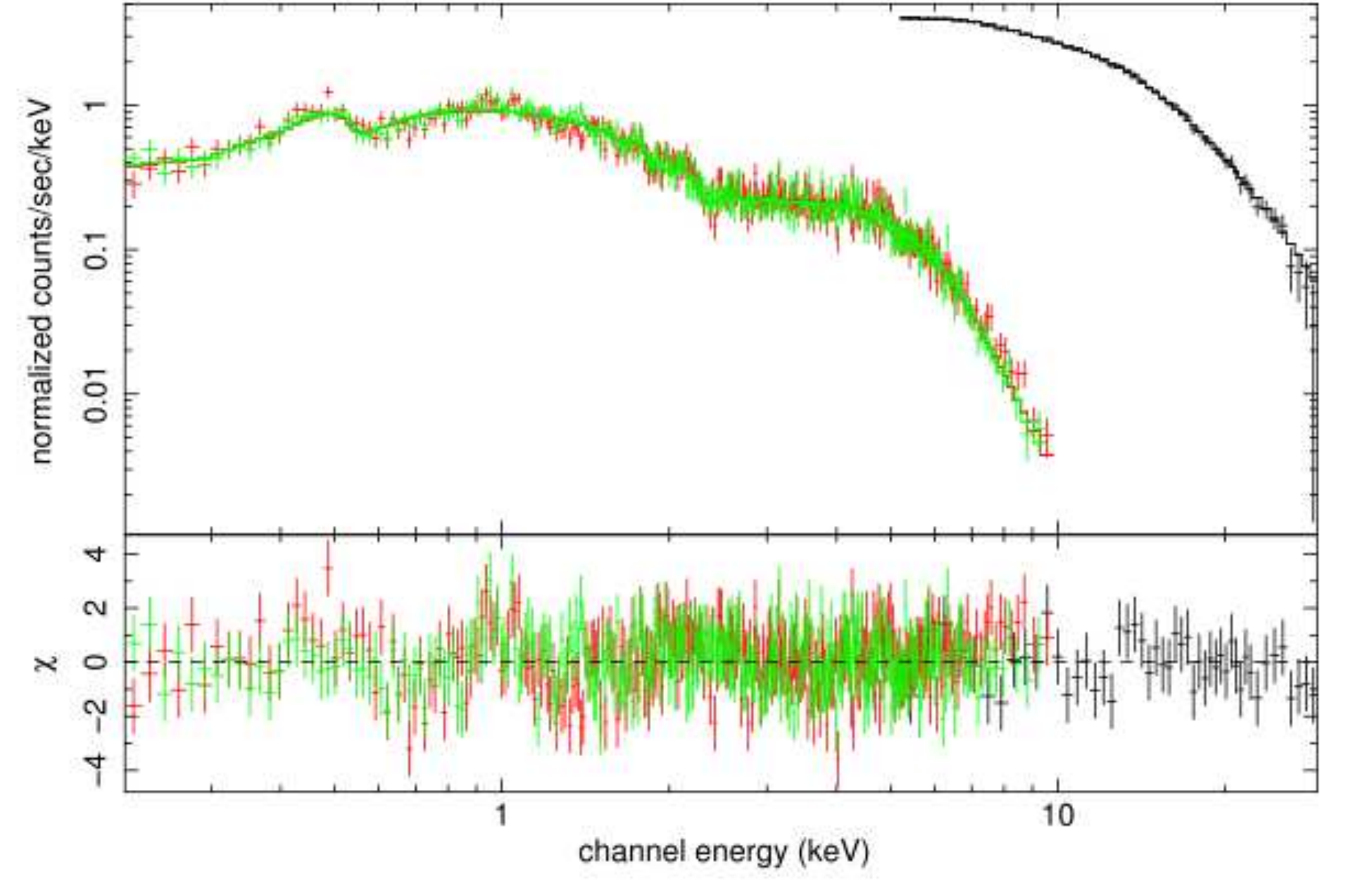}
\caption{The outburst spectrum is represented using a model invoking photoelectric absorption, a high-energy cutoff powerlaw, 
disk black-body, a single blackbody, and cyclotron-absorbing lines.We used EPIC/MOS1+2 and RXTE/PCA data.}
\label{fig:sepc:ds2}
\end{figure}

We finally added a sixth model component, a cyclotron line to account for the residuals observed around $\sim 10$ keV.
This model (\verb=const*wabs*(cutoff*cyclabs+diskbb+bb)=) fits the data very well (fig. \ref{fig:sepc:ds2})  providing {\bf $\chi_{\nu}^2 = 1.04$}, and 
a cross-calibration factor $\sim 0.9$. Table \ref{table:fit_diskbb_bb} lists the best fit parameters. The continuum, characterised by $\Gamma=0.2$ and E$_{C} =12$ keV, is typical for an accreting pulsar. 
The energy of the fundamental line ($9.0\pm 0.4$ keV) is   
consistent with the value derived from the RXTE data alone \citep{Markwardt_et_al_07}. 
We do not find evidence for cyclotron harmonic lines. 

\begin{figure}
\centering
\includegraphics[width=0.35\textwidth,angle=270]{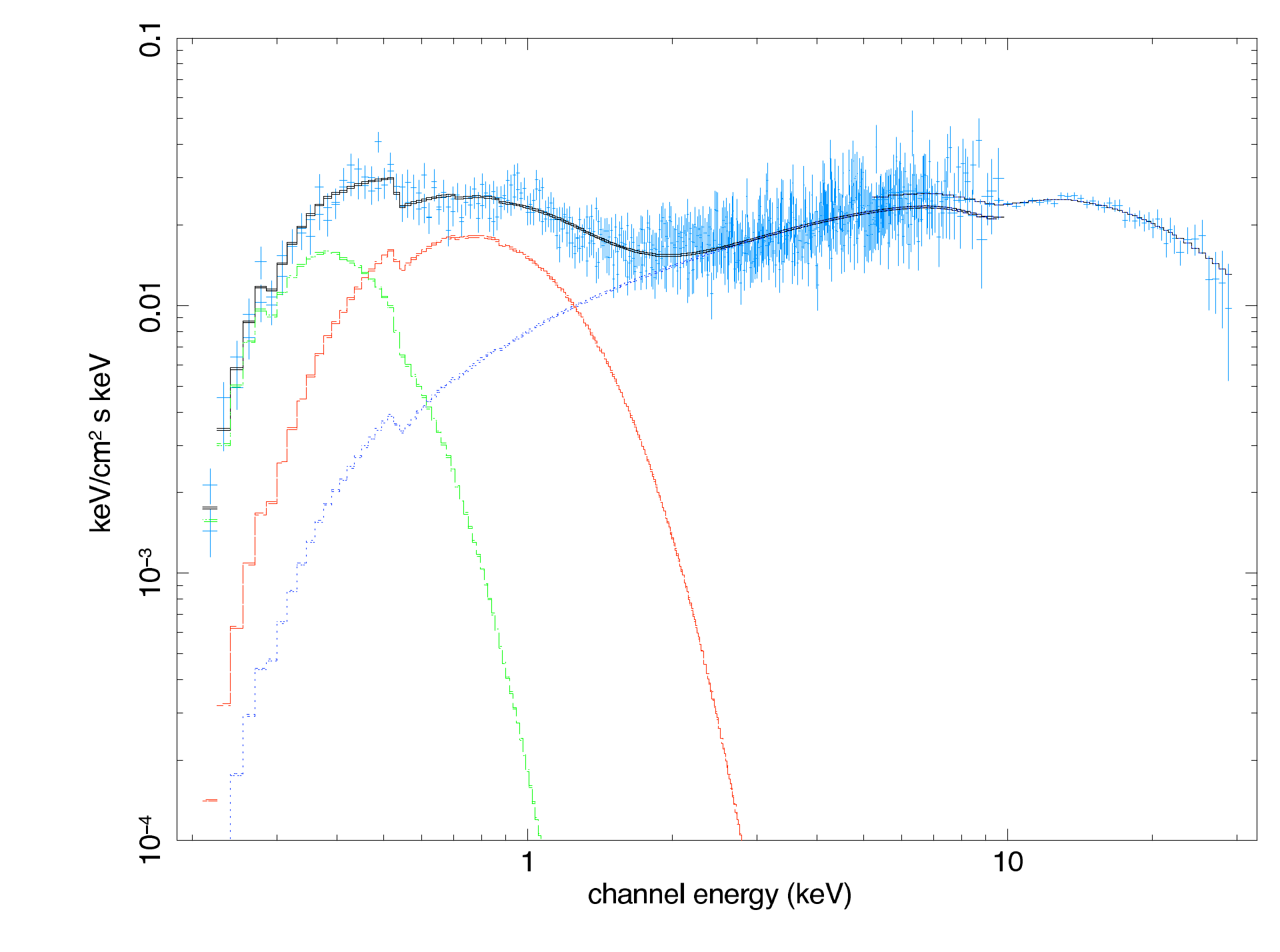}
\caption{$E\times f(E)$ outburst spectrum. The black solid line is the model described in the text. Cyan points are the unfolded spectrum. Blue, red, green dotted lines represent the cutoff powerlaw, hot blackbody and disk emission, respectively. For a color version see the electronic version of the paper.}
\label{fig:unfolded_outburst}
\end{figure}

Figure \ref{fig:unfolded_outburst} shows the $E\times f(E)$ model, unfolded spectrum, and the additive components of the model. The 
relative contribution of each component is also plotted. Above 5 keV the powerlaw cutoff component (blue dotted line) dominates the spectrum. 
At low energies, the blackbody (red dotted line) and the disk emission (green dotted line) dominates at 1 keV and below 0.5 keV, respectively. 

Based on this model, we estimated the unabsorbed flux for each component in the energy range $0.2 - 30$ keV as,
${\rm L_{0.2-30 keV}^{PL}\sim 3 \times 10^{38} \, erg\, s^{-1}}$, ${\rm L_{0.2-30 keV}^{BB}\sim  10^{37}\, erg\, s^{-1}}$, ${\rm L_{0.2-30 keV}^{DBB}\sim 10^{37}\, erg\, s^{-1}}$ 
given the distance to the LMC.

The residuals show some significant broad emission feature at 1 keV. Such features,
possibly an Fe L line complex \citep{Oosterbroek_et_al_97,McCray_et_al_82}, have been detected in other  X-ray binary pulsars \citep{Endo_et_al_00}.

\begin{table}
\begin{minipage}[t]{\columnwidth}
\caption{Best-fit parameters of the outburst spectrum (observation 2) for model: $const*wabs*(cutoff*cyclabs+diskbb+bbody)$.
Errors are calculated at $90\%$ confidence level independently for each parameter.} 
\centering
\renewcommand{\footnoterule}{}  
\begin{tabular}{lcc}
           Parameters    &   Value             &    Unit \\
\hline \hline
          $N_{H}$                   & $6.32\times10^{-2}\quad$\emph{(fixed)}              & $10^{22}$ cm$^{-2}$  \\
          $kT_{diskbb}$             & $ 0.088\pm 0.001$                             & keV     \\
          R$^{2}_{diskBB}cos\theta$ & $70000\quad$\emph{(fixed)}                          & km$^{2}$  \\
          $kT_{bbody}$              & $ 0.22\pm 0.01$                               & keV     \\
           $Area$                    &  $( 4.6\pm 0.1)\times 10^{4}  $             & km$^{2}$  \\
          $\Gamma$                  & $0.2\pm 0.1$                                 &            \\
          $E_{cutoff}$              & $ 12\pm 1$                                    & keV  \\
          Norm$_{cutoff}$           & $(1.09^{+0.06}_{-0.1})\times10^{-2}$         & ph keV$^{-1}$ cm$^{-2}$ s$^{-1}$ at 1keV    \\
          $E_{0}$                   & $9.0\pm0.4$                                  & keV \\
          $W_{0}$                   & $2.1^{+0.7}_{-0.6}$                          & keV  \\
          $\tau_{0}$                & $0.17 \pm 0.03$                              &        \\
          $C_{MOS1}$                & $0.90\pm 0.03$                               &        \\
          $C_{MOS2}$                & $0.89\pm 0.03$                               &         \\
          $\chi_{\nu}^{2}$/d.o.f    & $ 1.04/666$                                   &       \\

\hline
\end{tabular}
\label{table:fit_diskbb_bb}
\end{minipage}
\end{table}

\begin{table}
\begin{minipage}[t]{\columnwidth}
\caption{To search for spectral variability, all datasets were fit with a simple model (const*wabs*(cutoff+diskbb)). The normalization of the thermal and non-thermal components are listed below.
$kT_{in}=0.25$ keV, $\Gamma=0.42$, and $E_{C}=14$ keV, C$_{{\rm MOS1}}=0.89$, and C$_{{\rm MOS1}}=0.88$ are fixed. }
\centering
\renewcommand{\footnoterule}{}  
\begin{tabular}{lcccccc}
Obs.         & Normalization             &  Normalization                 & $\frac{F_{0.2-1}(BB)}{F_{0.2-1}(PL)}$ &  $\chi_{\nu}^{2}$    \\
             & \verb=diskbb=             & \verb=cutoff= $\times 10^{-3}$ &                                       &    (d.o.f)           \\
\hline \hline
1            & $53\pm4$                  &  2.51$\pm$0.05                 & 1.0 $\pm 0.1$                       & $1.20 (445)$       \\
2            & $890\pm20$                &  13.7$\pm$0.1                   & 3.23 $\pm 0.1 $                       & $1.26 (675)$       \\
3            & $58\pm4$                  &  2.77$\pm$0.05                 & 1.04  $\pm 0.1$                   & $1.34 (614)$       \\ 
4            & $2\pm1$                   &  0.23$\pm$0.01                   & 0.4  $\pm 0.2$                    & $1.05 (147)$       \\
\hline
\end{tabular}
\label{table:fit_pre_post}
\end{minipage}
\end{table}

\subsection{Spectral variability} 

To search for spectral variability during and outside of the outburst, we used a simple model to fit and compare 
the data of high and low signal to noise. 
To achieve this we removed the cyclotron line and the single black body component. 
The resulting model (\verb=const*wabs*(diskbb+cutoff)=) was first fit to the outburst data with all parameters free except $N_{H}$, fixed to the galactic value.
For the pre- and post-outburst  data several parameters were fixed 
to those obtained for the outburst spectrum: 
kT$_{in} = 0.25$ keV, $\Gamma=0.42$, and E$_{C}=14$ keV, C$_{{\rm MOS1}} = 0.89$, and C$_{{\rm MOS2}} = 0.88$. 
The remaining free parameters are the normalization 
of the disk blackbody, and of the cut-off power law. 
Table \ref{table:fit_pre_post} lists the results of our fits including the corresponging $\chi_{\nu}^{2}$ and the relative strength of the soft and hard components in the $0.2-1$ keV energy band.
The latter  peaked to a value of 3.2 for observation 2, during the outburst. 
The ratio was about 1.0 in observation 1 and 3 obtained close to the beginning and end of the outburst. Later
on, during observation 4, this relative strength decreased to 0.4. The outburst is characterized by a global
softening of the spectrum, signature of more efficient accretion.

\subsection{Timing analysis during the outburst}

\begin{figure}
\centering
$\quad \quad$ 
\vspace{-0.3cm}
\includegraphics[width=0.4\textwidth]{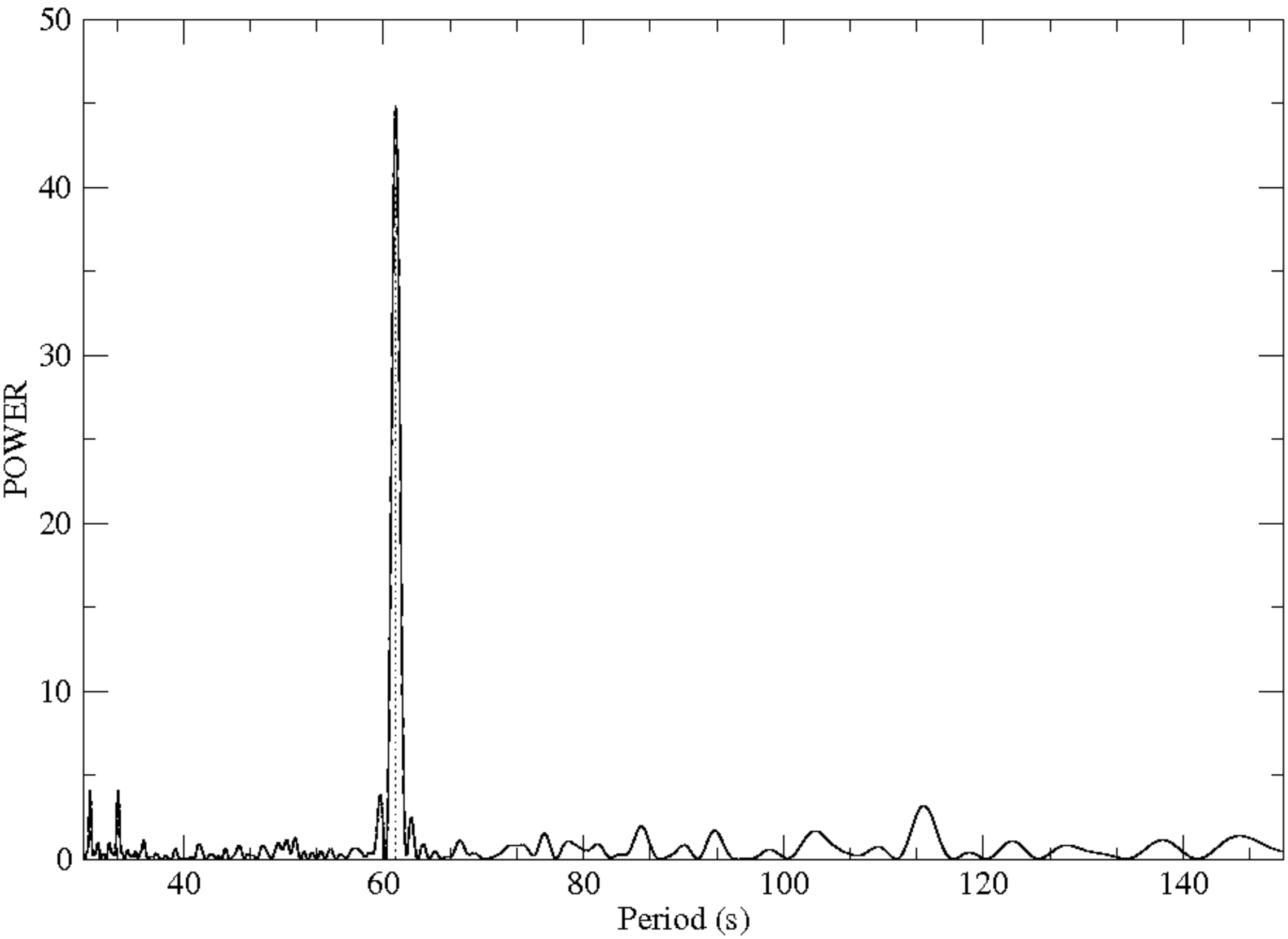}
\includegraphics[width=0.35\textwidth,angle=270]{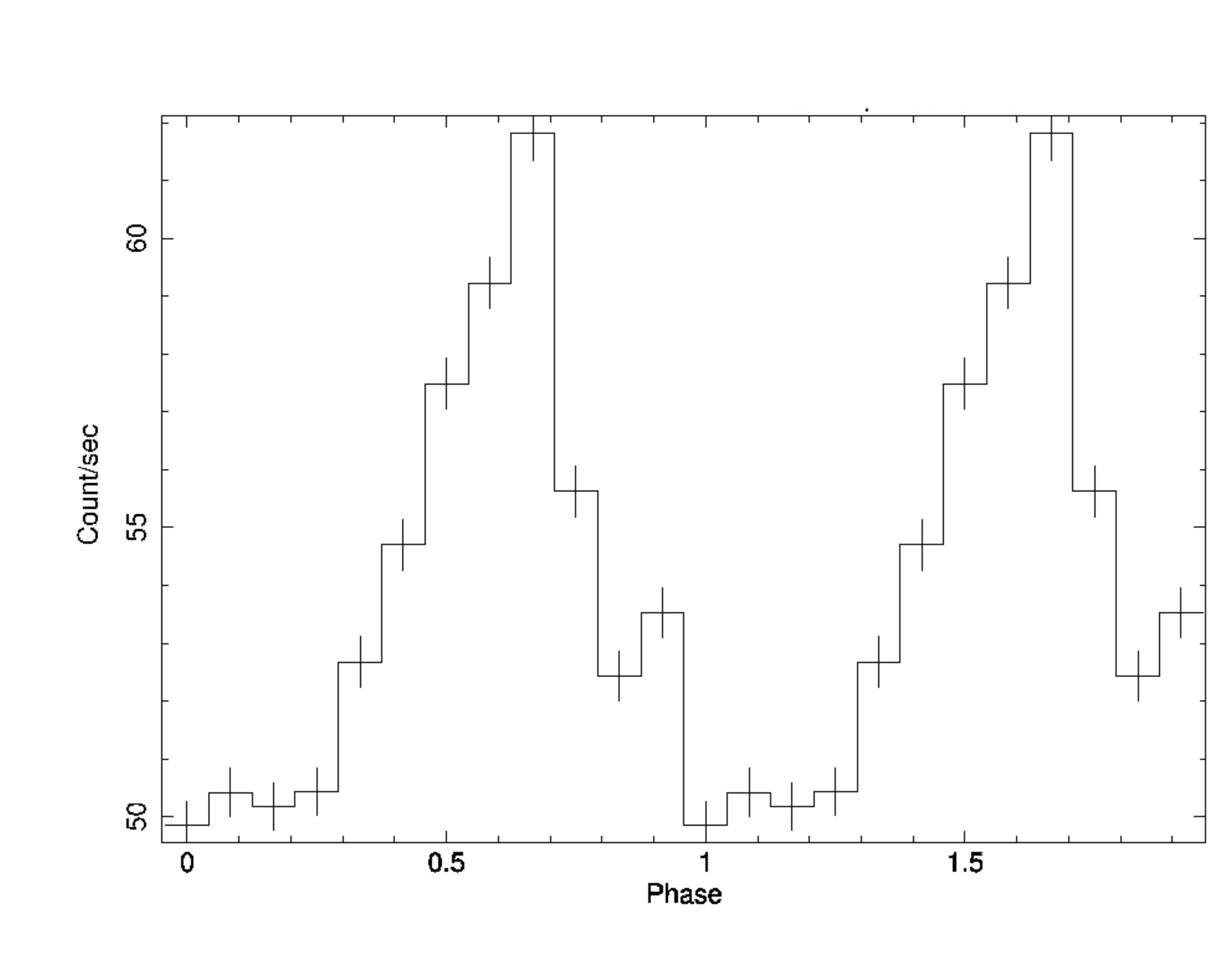}
\caption{\emph{Top:} Lomb-Scargle periodogram for observation 2 using RXTE $5-30$ keV data in standard-2 mode. The peak of the power spectrum 
density distribution is at ${\rm P}\approx 61.23 \pm 0.06$ s.
\emph{Bottom:} Folded lightcurve in the $5-30$ keV energy band obtained with this period. The zero epoch was set to MJD 54333.79882 }
\label{fig:p_rxte:ds1}
\end{figure}

We produced lightcurves in the energy bands $5-30$ keV (from RXTE standard-2 mode, i.e 16 seconds resolution),  $0.2-1$ keV and $3-10$ keV 
(from EPIC-MOS[1] event lists). Figure \ref{fig:p_rxte:ds1} shows the Lomb-Scargle periodogram, derived from the $5-30$ keV lightcurve,
with the power spectrum density distribution peaking at a period of  $61.23\pm 0.06$ s, and the corresponding folded lightcurve. 

In order to derive the likely pulse shape for the thermal spectral components, we have corrected the soft lightcurve for the contamination of the 
hard component in the soft band using $LC^{BB}_{{\rm soft}}=LC_{{\rm soft}}-\frac{C_{S}}{C_{H}}LC_{{\rm Hard}}$, where $\frac{C_{S}}{C_{H}}$
is the ratio of the counts predicted by the best fit model for both the soft and hard components, respectively. 

\begin{figure}
\centering
\includegraphics[width=0.45\textwidth]{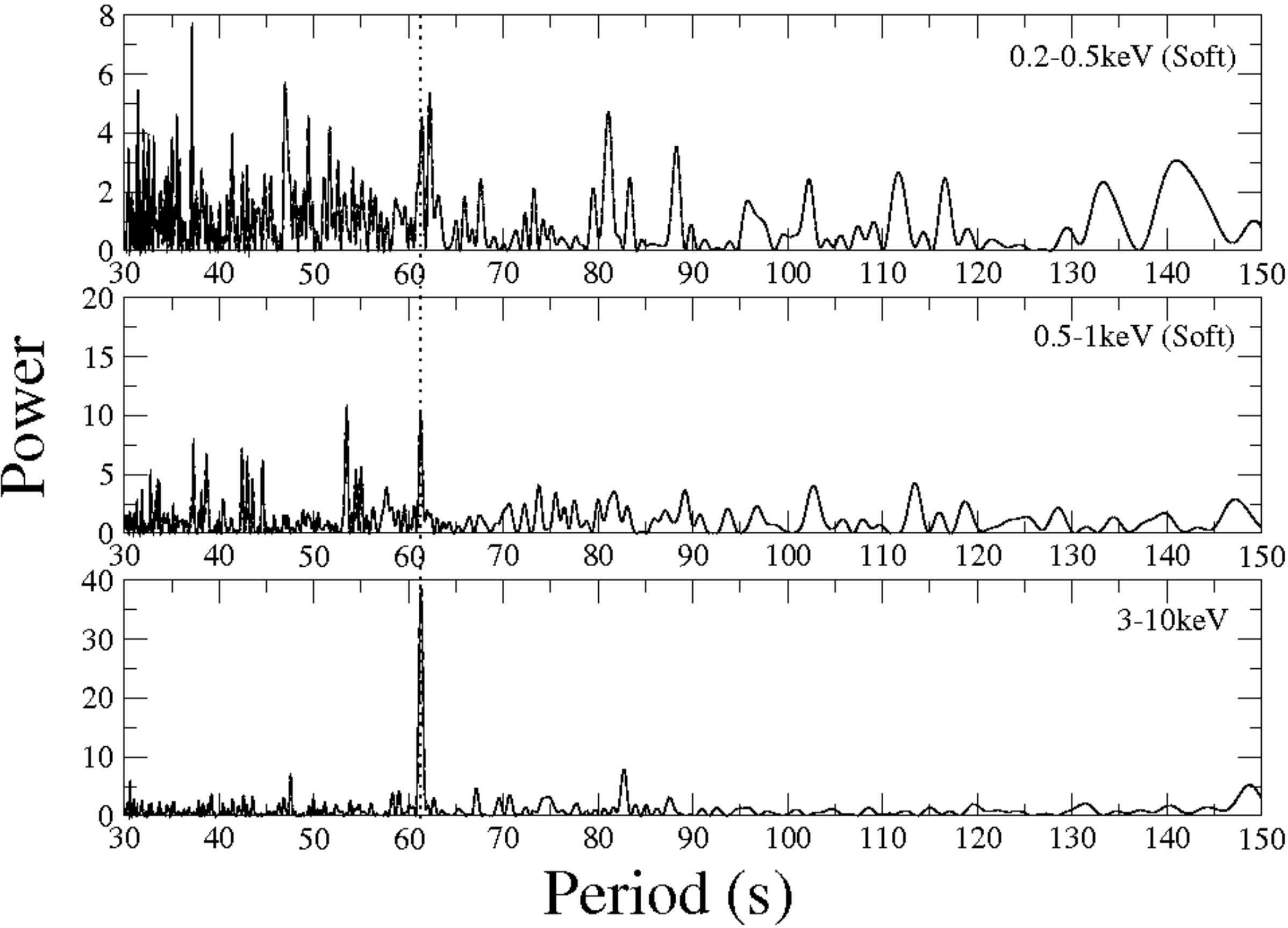}
\caption{Lomb-Scargle periodograms for observation 2 using XMM (EPIC/MOS[1]) data. The period  of 61.23 s is indicated with a dashed line. The first two soft X-ray periodograms were constructed
from lightcurves corrected for the contamination of the cutoff powerlaw component.}
\label{fig:p_xmm:ds1}
\end{figure}

\begin{figure}
\centering
\includegraphics[width=0.33\textwidth,angle=270]{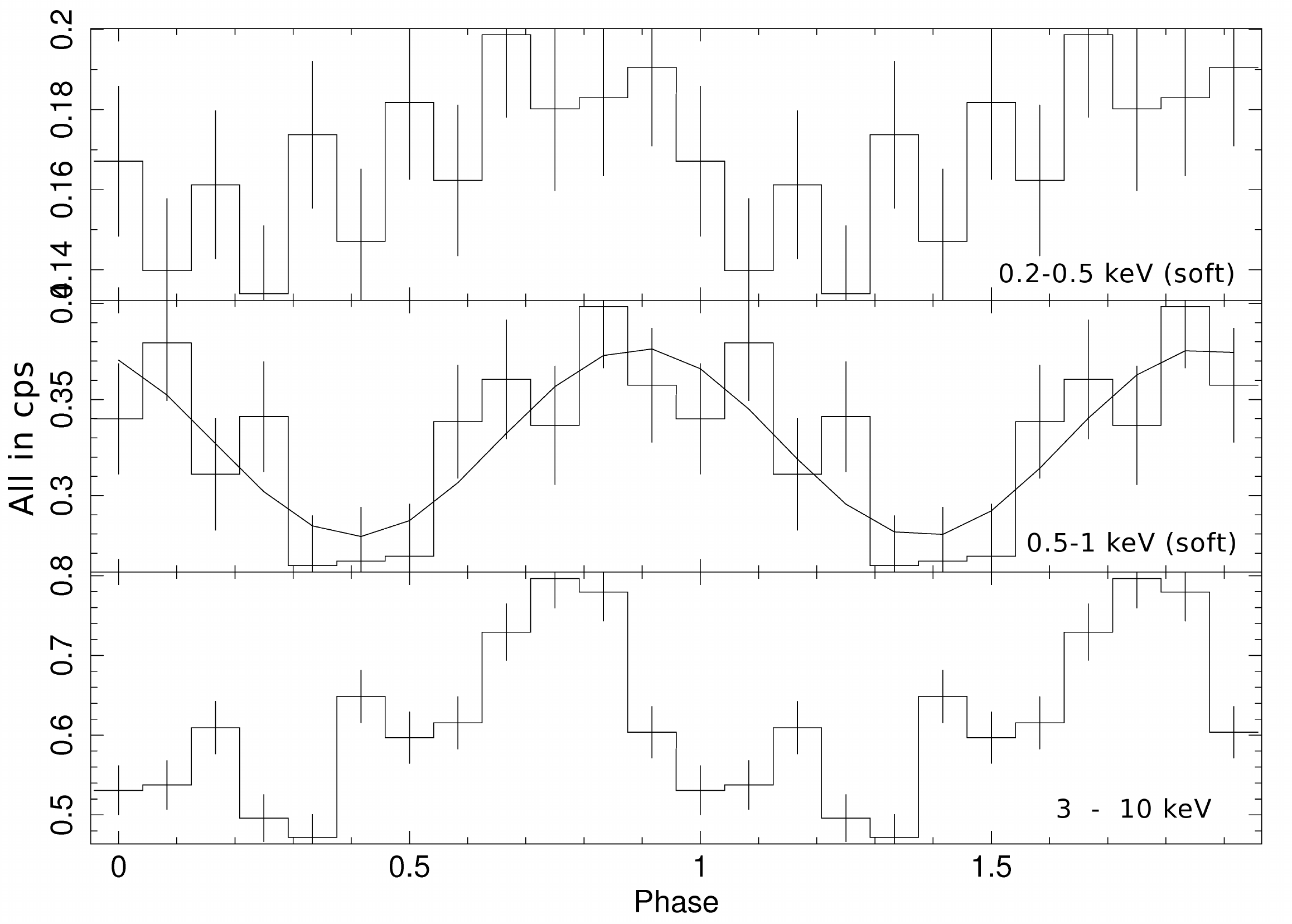}
\caption{Folded lightcurves obtained from EPIC/MOS[1] in the $0.2-0.5$ keV (top),  $0.5-1.0$ keV (middle), and $3-10$ keV (bottom)  energy bands, using 
the best spin period ($P\approx 61.23$s) found from RXTE/PCA data. The two soft lightcurves are corrected for the contamination of the cutoff powerlaw component. 
The solid curve in the middle panel shows a sinusoidal fit to the data. The zero epoch is set to MJD 54333.5}
\label{fig:p_xmm_fold_lc}
\end{figure}

Figure \ref{fig:p_xmm:ds1} shows the Lomb-Scargle periodograms obtained for the corrected ``soft'' $0.2-0.5$ keV, $0.5-1$ keV, and for the ``hard'' $3-10$ keV lightcurves. 
The corresponding folded lightcurves, using a period of ${\rm P}\approx 61.23$ s, are displayed in figure \ref{fig:p_xmm_fold_lc}.

The power spectrum density distribution at P=$61.23$s increases with energy (Fig \ref{fig:p_xmm:ds1}). 
The pulsation is well detected above $0.5$ keV and not significantly detected below $0.5$ keV. 

The shape of the peak in the folded $3-10$ keV lightcurve is sharper and narrower than in the ``soft'' lightcurves (Fig. \ref{fig:p_xmm_fold_lc}). The pulse starts 
simultaneously below 1 keV and above 3 keV, however the soft pulse appears to last twice the time.

\subsection{Variability of the spin period}

Table \ref{table:ls_periods} lists the periods derived for each observation in two energy bands, i.e. $0.2-10$ keV (obtained from XMM/EPIC-MOS[1]) and $5-30$ keV
(obtained from RXTE/PCA). Between observation 1 and 3, the neutron star spinned up by ${\rm \Delta P}=1.54$ s over a period of ${\rm \Delta T} \sim 50$ days.
Assuming a massive OB star ${\rm M \sim 20 \, M}_{\sun}$ and circular orbit with a period of 80 days the Doppler effects can be neglected ($\sim 10^{-3}$).

\begin{table}
\begin{minipage}[t]{\columnwidth}
\caption{The spin periods (in seconds) obtained from Lomb-Scargle periodograms, using data from \xmm and \rxte.}
\centering
\begin{tabular}{lcc}
Observation &     $0.2-10$keV        &  $5-30$keV       \\
\hline \hline
1           &     $62.19\pm0.02$     &       -         \\
2           &     $61.28\pm0.02$     &  $61.23\pm0.06$  \\
3           &     $60.65\pm0.01$     &  $60.62\pm0.3$  \\
4           &     $60.64\pm0.01$     &      -          \\
\hline
\end{tabular}
\label{table:ls_periods}
\end{minipage}
\end{table}

\subsection{Likely nearIR candidate}

The SWIFT XRT position provided by \cite{Palmer_et_al_07} is compatible with the position we derive
from the XMM-Newton MOS image (using SAS \verb=edetect= task): $\alpha_{2000}=5^{h} 41^{m} 34.33^{s},\,  \delta_{2000}=-68\degr 25\arcmin 49.0\arcsec$  (systematic uncertainty
of 2 arcsec). The XMM derived position is compatible with the optical counterpart detected by
SWIFT UVOT \citep{Palmer_et_al_07} corresponding to \xmmtwomass for which \citet{Curti_et_al_03} provide J=13.84$\pm$0.03 H=12.74$\pm$0.03, K=13.63$\pm$0.05  and
\citet{Monet_et_al_03} provide  B2=13.76, and  R2=13.84.

The infrared-optical spectral energy distribution of the likely stellar counterpart was fitted with a blackbody of  temperature $\sim 13000 \pm 1000~$K suggesting a B star. 
Additional reddening would increase the temperature. An infrared excess detected above the blackbody fit (in the H band) may be an instrumental effect or the signature of a circumstellar disk \citep{Wilson_et_al_05}.

\begin{figure}
\centering
\includegraphics[width=0.45\textwidth]{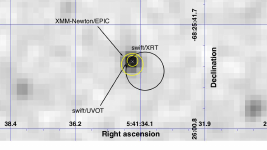}
\caption{K-band image ($49''\times 26''$) around \xmmu$\,$ obtained using the  2MASS/IPAC image server.
The circles indicate the source localization obtained by SWIFT/XRT (uncertainty 3.5$''$), and XMM/EPIC-MOS[1] (see text). The SWIFT/UVOT (uncertainty 1$''$) localization of the source
is consistent with \xmmtwomass.}
\label{fig:2mass_coordin}
\end{figure}


\section{Discussion}

The fundamental cyclotron absorption line, located at $\sim 9$ keV, indicates a magnetic field of ${\rm B}=8.6(1+z) E \times 10^{10}\, {\rm G} \approx 10^{12}\, {\rm G}$ (where $z=0.3$ is the gravitational redshift). The magnetospheric radius  ($\sim 1.3 \times 10^{8}\, {\rm cm}$) can be obtained 
by equating the kinetic energy density of the accreted material to the magnetic energy density \citep{White_Stella_88}.

The soft X-ray excess was modeled with a cold disk and a hot blackbody. 
The inner radius of the disk component was fixed to the radius of the  magnetosphere. 
The surface emitting the hot blackbody component can be estimated from the spectral fit as 4.6$ \times 10^{4}$ km$^{2}$.
Assuming that this corresponds to a broadened inner disk, its thickness can be estimated as $h=(4\pi {\rm R_{BB}}^{2})/(2\pi {\rm R_{m}}) \approx$  75 km.

If the inner disk is heated by the pulsar's hard X-ray emission, the covering factor can be estimated as $\frac{F_{BB}}{F_{PL}} \sim 0.03$, corresponding to $h\approx 45$ km.
The hot blackbody could therefore indeed be produced by reproccessing of hard x-rays on a broadened accretion disk.

Spectral variability was studied by applying a simple model to all the datasets. 
During the outburst the thermal component is more enhanced than the powerlaw component, suggesting a more effective accretion ({\bf $\dot{\rm M}/ \dot{\rm M}_{\rm Edd} \approx$ {\rm 0.8}}) and/or a broadened accretion disk.

The cooling time scale of the hot blackbody component ($\sim 10^{-6}$ s) is much shorter than the pulse 
period. When the neutron star rotates, the maximum of the reprocessed hard X-ray emission will move along the inner edge of the accretion disk. Pulsations of the soft blackbody are therefore expected.

\begin{figure}
\includegraphics[angle=270,bb=150 70  345 842,width=10cm,clip]{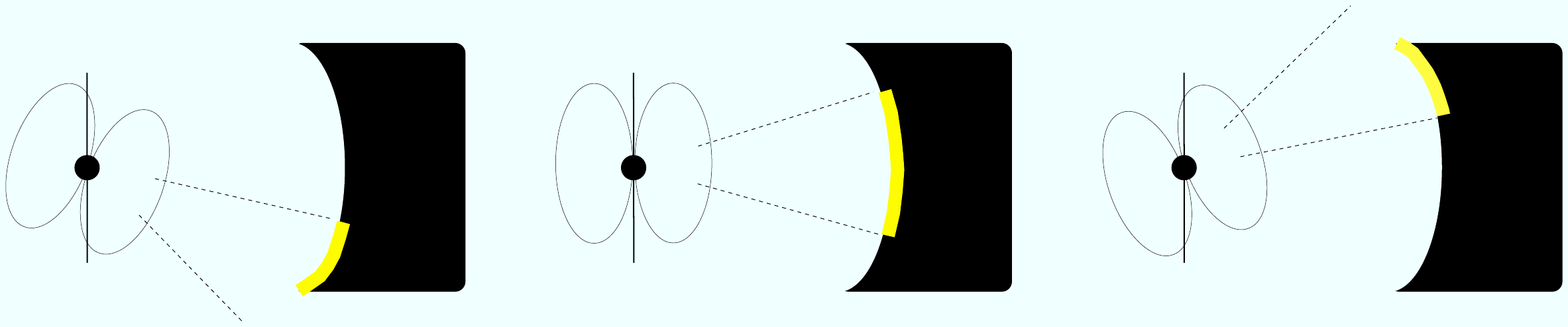}
\caption{Schematic representation of the reprocessing geometry in the accreting pulsar \xmmu. 
The distance between the neutron star and the disk is not to scale. 
The vertical line represents the rotation axis of the neutron star.
The yellow (grey) stripe indicates the reprocessing area on the inner edge of the accretion disk. The hard X-rays beam is indicated by dashed lines. 
 \emph{Left}: Soft X-ray minimum. \emph{Center}: Soft X-ray maximum. \emph{Right}: Soft X-ray minimum. }
\label{fig:reprocessing_geometry}
\end{figure}

Figure \ref{fig:p_xmm_fold_lc} shows the different shapes of the folded lightcurves in the soft and hard energy bands (0.5-1 keV and 3-10 keV). The ``hard'' pulse ($\Delta \phi \sim 0.4\pm0.1$) is shorter than the the hot blackbody pulse ($\Delta \phi \sim 0.8\pm0.1$).  The folded lightcurve of the hot blackbody component can in fact be represented by a sine curve (continuous line in the middle panel of fig. \ref{fig:p_xmm_fold_lc}) which could be expected in the very simplified source geometrical model represented in figure \ref{fig:reprocessing_geometry}. In this model we assume that the hard X-rays are emitted preferentially towards the magnetic equator \citep{Becker_Wolff_07} of the neutron star and that these photons illuminate the inner edge of the accretion disk. The illuminated disk re-radiates at soft X-rays and features a sinusoidal modulation.

We have also observed a significant spin-up of the pulsar ($\Delta {\rm P}\sim 1.5$ s) over a period of $\sim 50$ days,
i.e. $\dot{{\rm P}}\sim 3.5\times 10^{-7}\sim 11 {\rm s/yr}$ and ${\rm \dot{P} /  P} \approx 6 \times 10^{-9}$ s$^{-1} \sim 0.18$ yr$^{-1}$. 

\citet{Ghosh_Lamb_1979_2, Ghosh_Lamb_1979_3} described the interaction between the accretion disk and the stellar magnetic field and calculated the effect of the accretion torque on the spin of the neutron star. 
A relation between the mass M and radius R of the neutron star can be estimated for a given magnetic field, luminosity, spin period, and period derivative 
observed during the outburst. Our observations indicate a relation M/$M_{\odot}$ $\approx$ R/10 km, which is compatible with a range of neutron star equations of state \citep{Lattimer_et_al_01}, and an accretion rate of $\sim 3\times 10^{-8}$ M$_{\odot}\, yr^{-1}$. 

\xmmtwomass is the likely optical counterpart of \xmmu. The available photometry suggests a temperature of $13000\pm 1000$ K, confirming a B stellar counterpart.  

\section{Conclusion}

We report on XMM-Newton and RXTE observations of the X-ray binary pulsar \xmmu, performed in August 2007, during a giant type II outburst lasting for roughly 50 days. 
The outburst spectrum was fit successfully with a power law modified by an exponentional high energy cutoff, a cyclotron absorption line, and two soft thermal components.
We summarize our results as follows:
\begin{itemize}
\item  The reprocessing region corresponds  to the broadened inner edge of the accretion disk  broadened to $\sim$  75 km. 
\item The soft X-ray pulse shape profile (0.5-1 keV) shows sinusoidal modulation, a signature of illumination of the broadened inner disk.
\item The spin up of the pulsar and the enhancement of the disk emission during the outburst  indicate a high accretion rate ($\dot{M}/\dot{M}_{EDD}\approx$  0.8).
\item The infrared-optical spectral energy distribution of the counterpart suggests a hot primary star of T$\sim$13000K, likely a B-type star.
\end{itemize}

%
\begin{acknowledgements}
M.A. acknowledges support from a Swiss National Science Foundation Professorship (PP002--110504).
T. Lanz was supported by NASA grant NNX07AQ47G
This research has made use of NASA's Astrophysics Data System Bibliographic Service. 
The ASM lightcurve was obtained from the quick-look results provided by  the ASM/RXTE team.
\end{acknowledgements}

\bibpunct{(}{)}{;}{a}{}{,} 
\bibliographystyle{aa} 
\bibliography{paper_refereces} 
\end{document}